\documentclass[conference]{IEEEtran}
\IEEEoverridecommandlockouts
\usepackage{cite}
\usepackage{amsmath,amssymb,amsfonts}
\usepackage{algorithmic}
\usepackage{graphicx}
\usepackage{textcomp}
\usepackage{url}
\usepackage{xcolor}
\def\BibTeX{{\rm B\kern-.05em{\sc i\kern-.025em b}\kern-.08em
    T\kern-.1667em\lower.7ex\hbox{E}\kern-.125emX}}
\usepackage{bm}
\usepackage{booktabs}
\usepackage{flushend}
\begin{document}

\title{What Do Neurons Listen To? A Neuron-level Dissection of a General-purpose Audio Model
  \thanks{This work was supported by JST SICORP Grant Number JPMJSC2306 and JSPS KAKENHI Grant Number 24KJ1866.}
}

\author{%
  \IEEEauthorblockN{Takao Kawamura, Daisuke Niizumi, and Nobutaka Ono}
  \IEEEauthorblockA{
    \textit{Graduate School of Systems Design, Tokyo Metropolitan University} \\
    Tokyo, Japan  \\
    kawamura-takao@ed.tmu.ac.jp,
    daisukelab.cs@gmail.com,
    onono@tmu.ac.jp
  }
}

\maketitle

\begin{abstract}
In this paper, we analyze the internal representations of a general-purpose audio self-supervised learning (SSL) model from a neuron-level perspective.
Despite their strong empirical performance as feature extractors, the internal mechanisms underlying the robust generalization of SSL audio models remain unclear.
Drawing on the framework of mechanistic interpretability, we identify and examine class-specific neurons by analyzing conditional activation patterns across diverse tasks.
Our analysis reveals that SSL models foster the emergence of class-specific neurons that provide extensive coverage across novel task classes. These neurons exhibit shared responses across different semantic categories and acoustic similarities, such as speech attributes and musical pitch. We also confirm that these neurons have a functional impact on classification performance. To our knowledge, this is the first systematic neuron-level analysis of a general-purpose audio SSL model, providing new insights into its internal representation.
\end{abstract}


\section{Introduction}
Self-supervised learning (SSL)-based general-purpose audio representation models~\cite{gong2022ssast,niizumi2024masked,Li2023ATST-TALSP} achieve strong performance across a wide range of tasks, including audio classification and captioning.
These models encode acoustic information into general-purpose representations that can be used for downstream tasks.
Their effectiveness has primarily been evaluated through downstream task performance.

However, how such robust generalization is manifested within the model's internal representations remains an open question.
While evaluations on diverse downstream tasks demonstrate multifaceted practical effectiveness, they provide limited insight into how acoustic information is encoded \emph{inside} the model.
A deeper understanding of how these internal representations emerge is therefore essential to explain the success of general-purpose audio foundation models.

In parallel, recent studies have increasingly focused on mechanistic interpretability to investigate the mechanisms underlying neural network generalization~\cite{sharkey2025open}. At the neuron level, particularly in large language models (LLMs), these approaches analyze conditional activation patterns to localize attribute-sensitive neurons, such as those associated with language~\cite{tang2024language} and culture~\cite{namazifard2025isolating}. Such analyses provide fundamental insights into how general-purpose models mechanistically encode heterogeneous information.

In this paper, we address the lack of interpretability in general audio foundation models by applying a neuron-level analysis framework based on~\cite{tang2024language} to a SSL-based general-purpose model~\cite{niizumi2024masked} that exhibits strong generalization performance. Centered on the identification of class-specific neurons, we conduct a comprehensive investigation into how these units behave across diverse tasks, spanning environmental, speech, and music audio.
Through an analysis across diverse tasks, we provide answers to our fundamental research questions: 
(1) Are there class-specific neurons in unseen tasks? (2) What do these neurons share across different
classes? and (3) Do class-specific neurons contribute to classification?

Our analysis reveals that SSL models develop class-specific neurons that provide near-complete coverage across diverse novel task classes. Notably, we identify shared neuronal responses for speech attributes (gender, language, arousal), musical pitch, and acoustic similarities (e.g., across music genres). The validity of these identified neurons is further supported by their measurable impact on classification performance. To our knowledge, this is the first systematic neuron-level analysis of a general-purpose audio SSL model.
We publicly release our code\footnote{The URL will replace this placeholder after the notification.} for further advances and reproducibility in the field.

\section{Identifying Class-specific Neurons}\label{sec:method}
We investigate whether audio models contain neurons that respond selectively to specific sound properties, and address this question through activation-based neuron analysis.
To this end, we identify class-specific neurons using Audio Activation Probability Entropy (AAPE), an entropy-based metric adapted from Language Activation Probability Entropy (LAPE)\cite{tang2024language} originally proposed in the NLP domain, which quantifies how selectively a neuron activates across sound classes.

For each layer $l$ and neuron $n$, AAPE is defined as
\begin{equation}
    \mathrm{AAPE}_{l,n}
    = - \sum_{c \in \mathcal{C}} \tilde{P}^{(c)}_{l,n}
    \log \tilde{P}^{(c)}_{l,n},
\end{equation}
where $\mathcal{C}$ denotes the set of event classes and
\begin{equation}
    \tilde{P}^{(c)}_{l,n}
    = \frac{P^{(c)}_{l,n}}{\sum_{c' \in \mathcal{C}} P^{(c')}_{l,n}} .
\end{equation}
Here, $P^{(c)}_{l,n}$ represents the \emph{class-wise activation probability}, defined as the proportion of samples belonging to class $c$ for which neuron $(l,n)$ produces a positive activation (i.e., activation value greater than zero).
A lower AAPE value indicates that the neuron tends to activate preferentially for a limited subset of sound classes, and such neurons are regarded as more class-specific.

To identify class-specific neurons, we follow LAPE by applying a three-step filter: (1) excluding the bottom 5\% of neurons with insufficient activation, (2) selecting those within the lowest $r_\text{AAPE}$ percentile of AAPE scores to ensure high selectivity, and (3) designating the top 5\% by class-specific activation probability as the target neurons. The threshold $r_\text{AAPE}$ is calibrated based on the task's class composition.

Given that our target model is based on the Transformer encoder architecture, we define the activations within the encoder blocks as individual neurons, a practice commonly adopted in the interpretability literature (e.g., LAPE).

\section{Empirical Analyses}
In the experiments, we identified neurons responsive to diverse sounds and analyzed their relationships across different classes and tasks (datasets) to better understand the mechanisms underlying the model's generalization. Additionally, we conducted ablation by steering the identified neurons' activity to observe its impact on classification performance.

\subsection{Experimental Setup}
We identified class-specific neurons for each task, represented by unique IDs $\langle \text{layer, neuron} \rangle$, to detect overlaps across classes and tasks. The target model consists of 12 layers, each containing $768 \times 4 = 3072$ neurons. We set the entropy threshold $r_\text{AAPE}$ to 2\% for ESC-50 and GISE-51 due to their diverse class compositions, while 1\% was used for other tasks following LAPE.

To evaluate classification performance during steering ablations, we employed linear evaluation using the evaluation platform (EVAR\footnote{\url{https://github.com/nttcslab/eval-audio-repr}}) following \cite{niizumi2024masked}.

\subsubsection{Models}
We utilized a Vision Transformer\cite{ViT} (ViT) pre-trained via Masked Modeling Duo\cite{niizumi2024masked} (M2D), a SOTA self-supervised model. M2D is ideal for our analysis because it learns robust audio representations through simple masked prediction, demonstrating high generalization\footnote{We used the M2D/0.7 weight file from \url{https://github.com/nttcslab/m2d}}. For comparison, we also evaluated a ViT with the same architecture, randomly initialized and trained via supervised learning (SL) on AudioSet\cite{gemmeke2017audioset} from scratch.
Hereafter, we denote the former as \emph{SSL ViT (M2D)} and the latter as \emph{SL ViT}.

\subsubsection{Tasks (Datasets)}

We employed diverse tasks mostly following \cite{niizumi2024masked}, focusing on single-source clips to ensure that our neuron-level analysis remained unaffected by multiple acoustic features.

\begin{itemize}
    \item ESC-50~\cite{piczak2015esc} comprises 2000 samples across 50 general sound event classes, including animal vocalizations, natural soundscapes, and urban noises such as sirens. Due to its class-balanced design, it serves as a controlled benchmark for evaluating our analysis.
    \item GISE-51~\cite{yadav2021gise} consists of 51 isolated general sound event classes with 16,357 samples. % curated from FSD50K.
    Although it shares a similar class taxonomy with ESC-50, GISE-51 presents a more challenging scenario due to its class imbalance and variable sample durations.
    \item VoxForge~\cite{voxforge} comprises speech samples across six language classes. We utilized 5000 samples per language, for a total of 30,000.
    \item VoxCeleb1~\cite{nagrani2017voxceleb} (VC1) comprises diverse speech from individuals. We selected 1500 samples for which both gender (male/female) and nationality (spanning 35 countries) are available.
    \item CREMA-D~\cite{cao2014crema} comprises 7438 largely class-balanced speech samples with acted emotion classes. Additionally, we utilized gender information as an alternative class in our experiments.
    \item GTZAN~\cite{tzanetakis2002musical,sturm2013gtzan} comprises 1000 class-balanced music audio samples across ten genres.
    \item Pitch Audio Dataset (Surge synthesizer)~\cite{turian2021one} (Surge) comprises synthesizer note samples. We categorized the 44,000 samples into 9 octave classes by aggregating the original 88 individual pitch classes.
    \item NSynth~\cite{engel2017neural} also consists of synthesizer note samples. For our experiments, we utilized a subset of 22,000 samples, labeled by either their 11 instrument family classes or 9 octave classes. We aggregated the original pitch classes into octave classes, as in Surge.
\end{itemize}

\subsection{RQ1: Are There Class-specific Neurons in Unseen Tasks?}
To address this question, we examine the emergence of class-specific neurons in various unseen tasks, providing insight into how models generalize beyond their training data.
Specifically, we compare the abundance and selectivity of these neurons in SSL and SL models to assess their capacity for representing novel categories.
Tables~\ref{tab:ssl-vit-total} and \ref{tab:sl-vit-total} summarize the statistics of class-specific neurons.

SSL achieves a class coverage ratio close to 100\% for most tasks, whereas SL covers substantially fewer classes (e.g., 49\% in VC1). SSL also yields twice as many mean class-specific neurons compared to SL.
Consequently, SSL fosters the emergence of a comprehensive set of class-specific neurons that cover novel class hierarchies, which likely underpins its robust generalization. In contrast, SL appears to be constrained by its fixed training class definitions, potentially hindering the development of neurons for unseen task categories and suggesting a fundamental bottleneck in supervised coverage.

\begin{table}[t]
\begin{center}
\caption{Statistics of Class-specific Neurons (SSL).
SSL achieves a class coverage ratio close to 100\%, indicating that class-specific neurons span nearly all classes.}
\label{tab:ssl-vit-total}
\begin{tabular}{lcc}
    \toprule
     & Mean class-specific neurons & Class coverage ratio\\ 
    \midrule
    CREMA-D & 172.0 & 100\%\\
    ESC-50 & 37.9 & 98\%\\
    GISE-51 & 42.3 & 100\%\\
    GTZAN & 70.2 & 100\%\\
    NSynth & 98.5 & 100\%\\
    Surge & 164.9 & 100\%\\
    VC1 & 28.7 & 100\%\\
    VoxForge & 199.2 & 100\%\\
    \midrule
    Average & 101.7 & 100\%\\
    \bottomrule
\end{tabular}
\end{center}
\end{table}

\begin{table}[t]
\begin{center}
\caption{Statistics of Class-specific Neurons (SL).
SL covers substantially fewer classes (e.g., 49\% in VC1), indicating that their neurons exhibit limited class-level coverage.}
\label{tab:sl-vit-total}
\begin{tabular}{lcc}
    \toprule
     & Mean class-specific neurons & Class coverage ratio\\
    \midrule
    CREMA-D & 98.7 & 100\%\\
    ESC50 & 24.7 & 94\%\\
    GISE-51 & 23.4 & 80\%\\
    GTZAN & 48.1 & 75\%\\
    NSynth & 49.7 & 100\%\\
    Surge & 81.7 & 100\%\\
    VC1 & 8.4 & 49\%\\
    VoxForge & 106.7 & 100\%\\
    \midrule
    Average & 55.2 & 87\%\\
    \bottomrule
\end{tabular}
\end{center}
\end{table}

\subsection{RQ2: What Do These Neurons Share Across Different Classes?}
In this section, we investigate the degree of sharing across classes and tasks among the neurons identified in SSL in the previous section.

\begin{figure}[t]
\centerline{\includegraphics[width=0.8\linewidth]{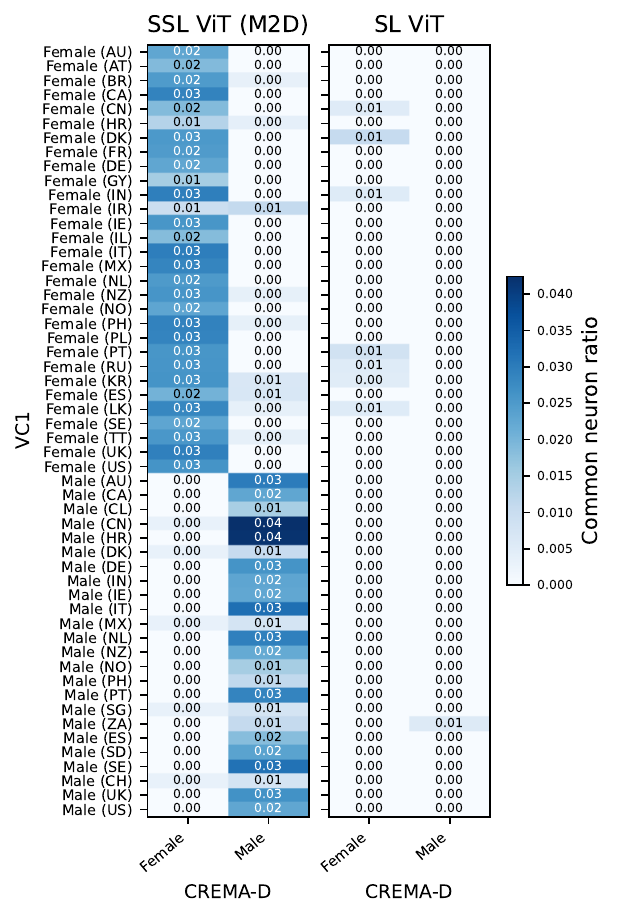}} % orig: overlap-vc1-cremad_gender_v2 (1).pdf
\vspace{-10pt}
\caption{Shared neurons between VC1 and CREMA-D under gender-based class definitions.
SSL demonstrates clear cross-task sharing of neurons aligned with gender, whereas SL exhibits negligible sharing.}
\label{fig-gender}
\end{figure}

Figure~\ref{fig-gender} shows the observations of shared neurons related to gender in speech.
For this analysis, classes in VC1 were defined by gender and nationality, whereas those in CREMA-D were defined solely by gender.
The values represent the Jaccard coefficient (intersection over union) between each class in VC1 and each class in CREMA-D, indicating the proportion of shared neurons.
In this paper, we refer to these values as \emph{common neuron ratios}.
The results show that neurons shared across tasks are substantially more numerous in SSL than in SL. In particular, VC1 female classes share neurons primarily with CREMA-D female classes, and a similar pattern is observed for male classes. Although SSL is trained without explicit gender information, the neurons in SSL respond to speech characteristics associated with gender.

\begin{figure}[t]
\centerline{\includegraphics[width=\linewidth]{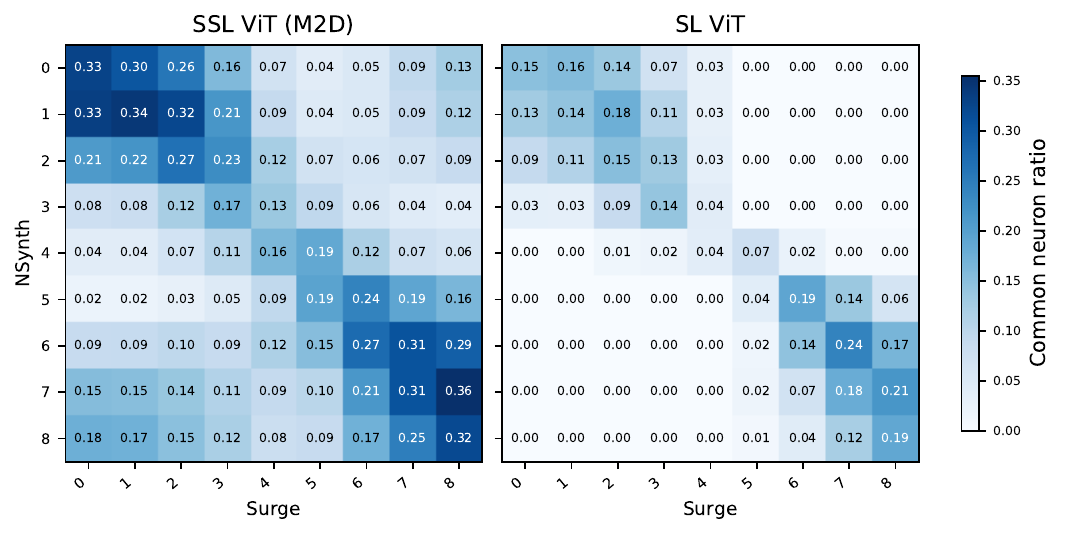}}
\vspace{-10pt}
\caption{Common neuron ratios for each octave class in NSynth and Surge.
Despite differences in synthesis methods and dataset characteristics, octave-specific neurons are consistently observed across both tasks.}
\label{fig-octave}
\end{figure}

Figure~\ref{fig-octave} presents the shared neurons associated with pitch in musical sounds.
NSynth and Surge consist of synthetic musical sounds generated using different synthesis methods.
For this analysis, nine octave-based labels were assigned to each task.
In contrast to the gender analysis (Fig.~\ref{fig-gender}), shared neurons associated with pitch are observed even in SL.
Moreover, in both SSL and SL, shared neurons are more frequently observed at both lower and higher octaves.
These results suggest that pitch height is a salient acoustic attribute that characterizes sounds.

\begin{figure}[t]
\centerline{\includegraphics[width=\linewidth]{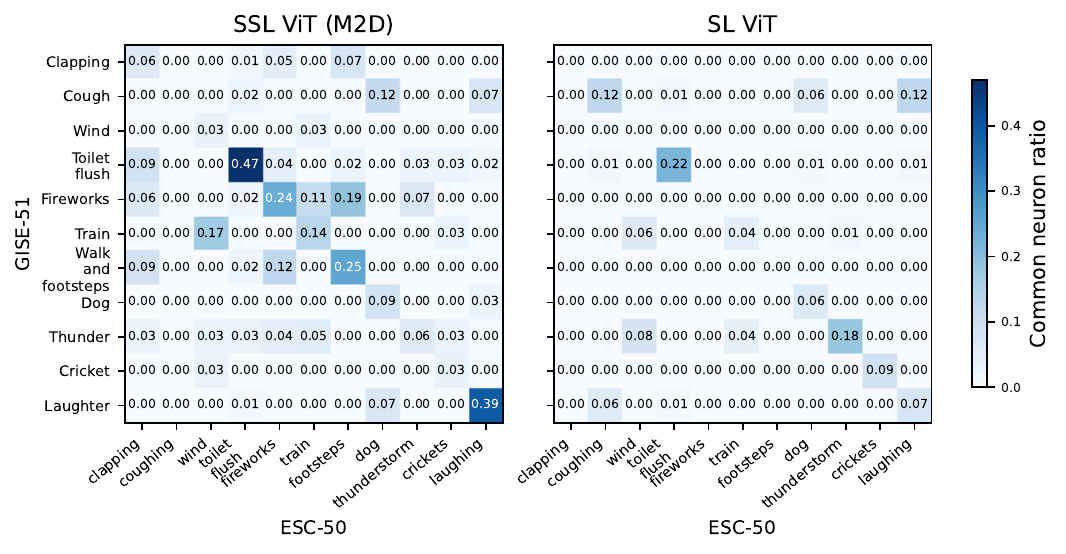}}
\vspace{-10pt}
\caption{Common neuron ratios for semantically overlapping event classes across ESC-50 and GISE-51.
Compared with Fig.~\ref{fig-gender} and Fig.~\ref{fig-octave}, the ratios are generally lower.}
\label{fig-event}
\end{figure}

Figure~\ref{fig-event} presents the shared neurons for semantically overlapping event classes across ESC-50 and GISE-51 (e.g., ``Cough'' vs. ``coughing'' and ``Thunder'' vs. ``thunderstorm'').
In contrast to the previous analyses (Fig.~\ref{fig-gender} and Fig.~\ref{fig-octave}), the common neuron ratios are generally lower overall.
The sounds within each event class appear to exhibit considerable variation, which may reduce acoustic commonality across datasets. 
Among these classes, ``Toilet flush'' and ``Laughter'' show higher common neuron ratios, suggesting more consistent acoustic patterns across datasets.
Conversely, for ``Train'' and ``Cough'', higher common neuron ratios are observed with acoustically different classes rather than with their semantically overlapping classes.
For example, ``Train'' exhibits higher commonality with ``Wind,'' likely reflecting the co-occurrence of background wind noise in train recordings. In contrast, ``Cough'' shows higher commonality with ``Dog,'' which may indicate acoustic similarity between certain cough sounds and dog vocalizations.

\begin{figure}[t]
\centerline{\includegraphics[width=\linewidth]{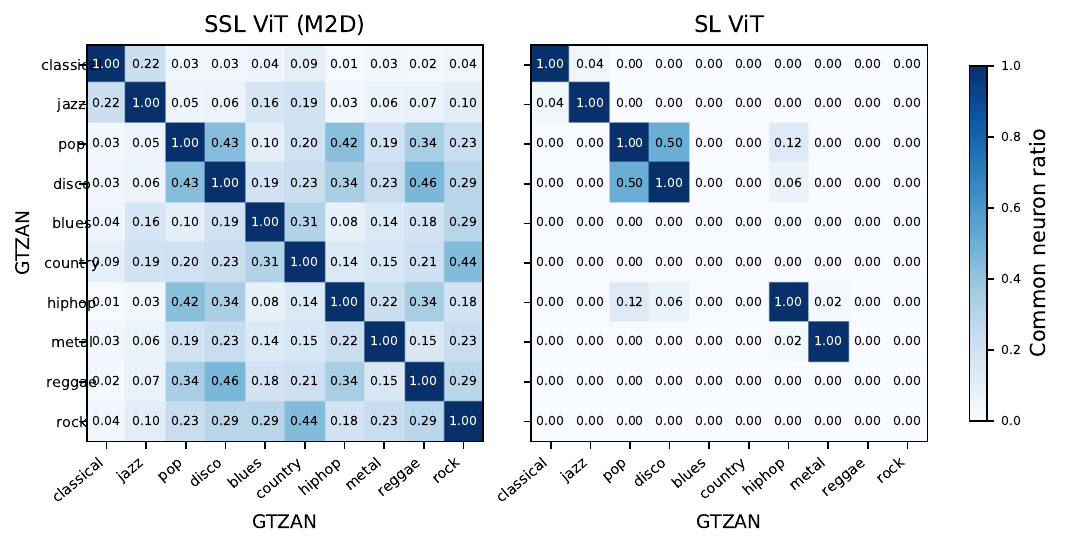}}
\vspace{-10pt}
\caption{Common neuron ratios across genre classes in GTZAN.
``Classical'' and ``jazz'' share a relatively large number of neurons, exhibiting a distinct sharing pattern compared with other genres.}
\label{fig-genre}
\vspace{-5pt}
\end{figure}

Figure~\ref{fig-genre} shows the sharing of neurons across genre classes in GTZAN.
In both SSL and SL, ``classical'' and ``jazz'' share more neurons than other genre pairs, indicating a distinct pattern of neuron sharing.
These results suggest that the observed neuron sharing reflects shared acoustic characteristics across these genres.
For example, ``classical'' and ``jazz'' often exhibit acoustic structures that differ from genres such as ``pop,'' where vocals are a primary component.
Such differences may contribute to the observed overlap in neuron activations.

\begin{figure}[t]
\centerline{\includegraphics[width=\linewidth]{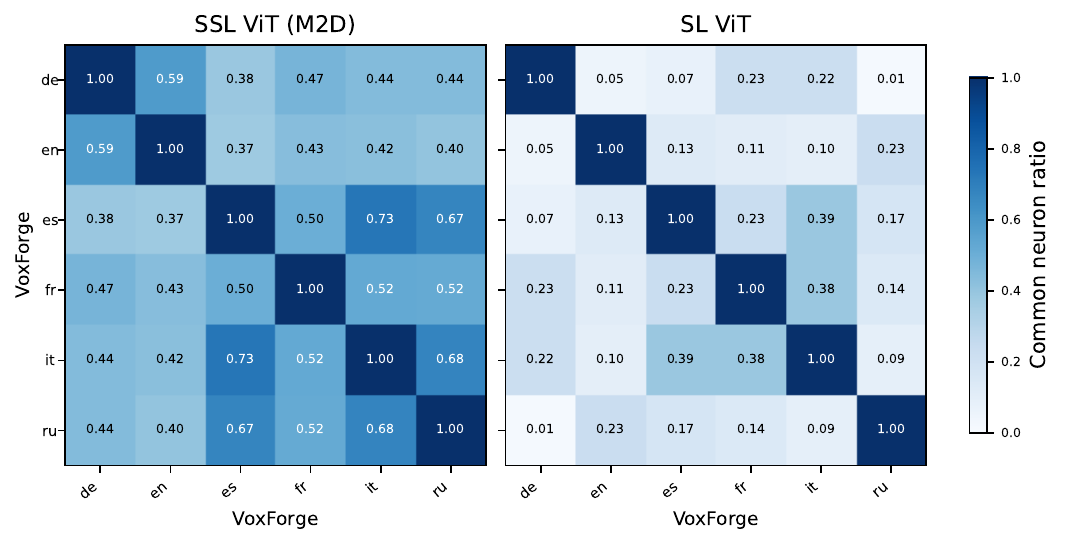}}
\vspace{-10pt}
\caption{Common neuron ratios across language classes in VoxForge.
SSL results exhibit relatively high neuron sharing within Germanic (``de,'' ``en'') and Romance (``es,'' ``fr,'' ``it'') language families.}
\label{fig-language}
\end{figure}

Figure~\ref{fig-language} presents the common neuron ratios across language classes in VoxForge.
In SSL, higher neuron sharing is observed between ``de'' and ``en,'' while ``es,'' ``fr,'' ``it,'' and ``ru'' also share relatively many neurons.
Germanic languages such as ``de'' and ``en'' are known for vowel reduction in unstressed syllables, whereas Romance languages such as ``fr,'' ``it,'' and ``es'' exhibit comparatively less vowel reduction and more stable vowel realizations~\cite{Fletcher2010Prosody}.
The SSL results suggest that such language-family-level phonetic differences may be reflected in the shared internal representations.

\begin{figure}[t]
\centerline{\includegraphics[width=\linewidth]{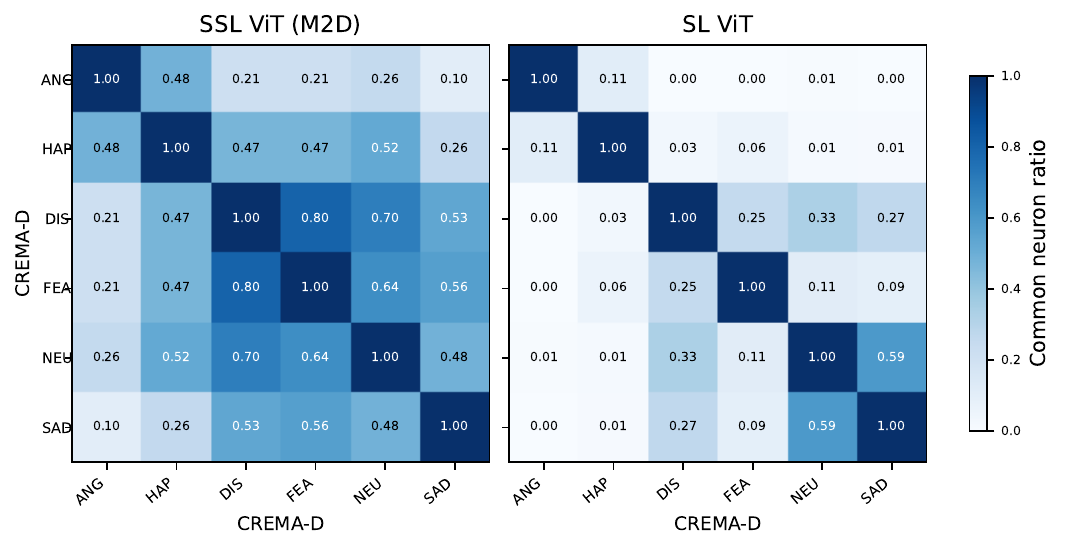}}
\vspace{-10pt}
\caption{Common neuron ratios across emotion classes in CREMA-D.
Relatively higher neuron sharing is observed between ``ANG'' and ``HAP,'' as well as among ``DIS,'' ``FEA,'' ``NEU,'' and ``SAD.''}
\label{fig-emotion}
\vspace{-5pt}
\end{figure}

Figure~\ref{fig-emotion} presents the common neuron ratios across emotion classes in CREMA-D.
``ANG'' and ``HAP'' show elevated neuron sharing, while relatively higher sharing is also observed among ``DIS,'' ``FEA,'' ``NEU,'' and ``SAD.''
This trend is partially aligned with an arousal-based organization of emotions.
In Russell’s circumplex model of affect~\cite{russell1989cross}, ``ANG'' and ``HAP'' are located in the high-arousal region, whereas ``SAD'' lies in the low-arousal region.
These findings suggest that acoustic cues related to arousal may be encoded in the learned representations.

In this section, shared neurons were observed across various tasks and classes, especially with SSL, suggesting that these neurons capture common acoustic characteristics across different tasks.
Furthermore, several results indicate that certain neurons are sensitive to such acoustic similarities.
These findings imply that SSL, through learning without explicit supervision, captures salient acoustic attributes that may support higher-level semantic classification tasks such as audio event recognition.

\subsection{RQ3: Do Class-specific Neurons Contribute to Classification?}

In this section, we conduct an ablation study to examine whether class-specific neurons contribute to the prediction of their corresponding classes.
We examine the effects of deactivating class-specific neurons on downstream classification performance by setting the outputs of their activation functions to zero during inference.

Specifically, we focus on GTZAN, targeting the 26 class-specific neurons shared between ``classical'' and ``jazz'' (as shown in Fig.~\ref{fig-genre}), and CREMA-D, targeting the 22 neurons shared between ``ANG'' and ``HAP'' (as shown in Fig.~\ref{fig-emotion}).
As a baseline for comparison, we randomly deactivate the same number of neurons.

\begin{figure}[t]
\vspace{-15pt}
\centerline{\includegraphics[width=\linewidth]{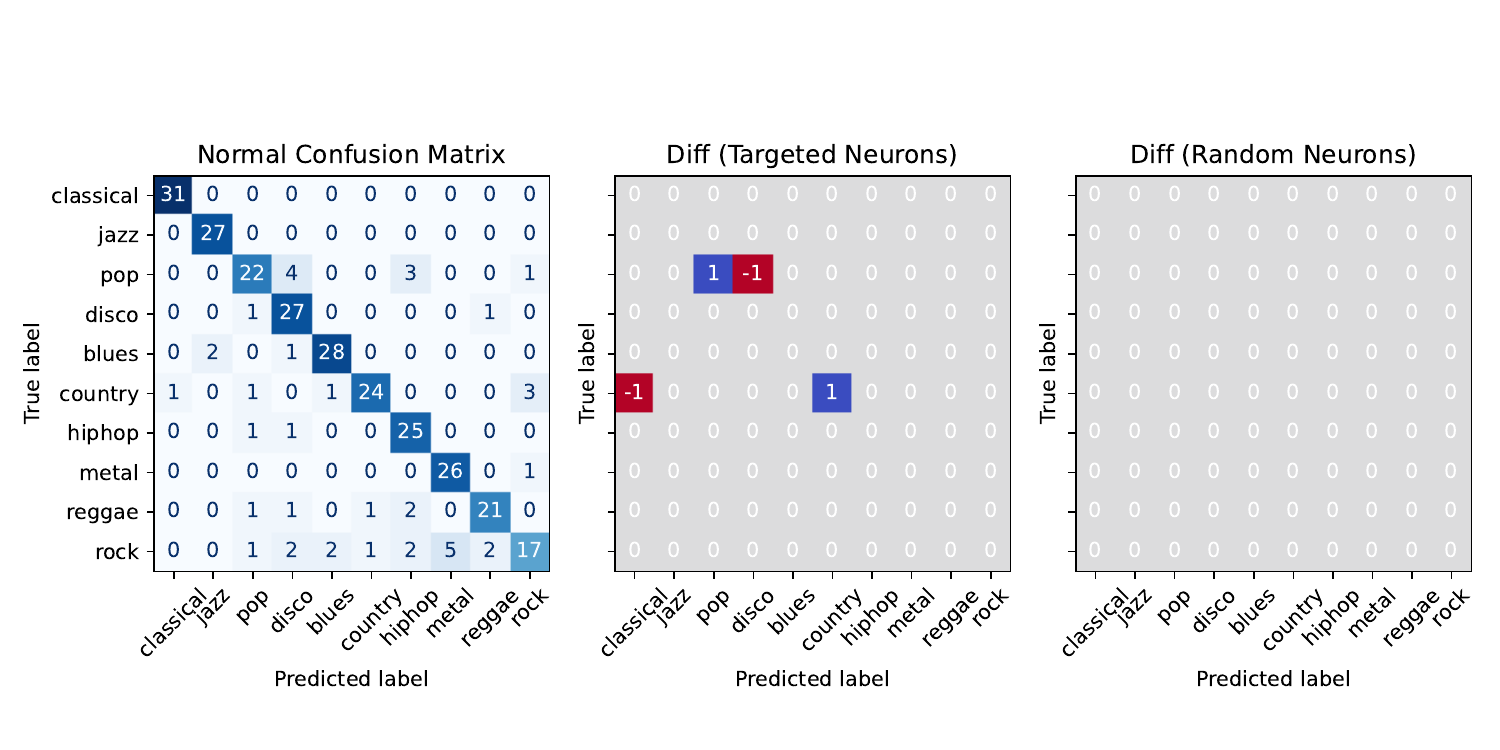}}
\vspace{-15pt}
\caption{Ablation impact on GTZAN genre classification. Left: Original model. Middle/Right: Deviations after ablating class-specific vs. random neurons. The results indicate that class-specific neurons have a greater functional impact on classification than randomly selected ones.}
\label{fig-cm-genre}
\end{figure}

The experimental results confirm that deactivating the class-specific neurons affects class-level classification performance.
In Fig.~\ref{fig-cm-genre} (GTZAN), although ``classical'' and ``jazz'' achieve 100\% accuracy in the original model, ablating the targeted neurons improves performance for ``country'' and ``pop,'' suggesting that reducing an over-reliance on these specific neurons mitigated previous misclassifications.
In contrast, random ablation had no impact on performance.

\begin{figure}[t]
\vspace{-10pt}
\centerline{\includegraphics[width=\linewidth]{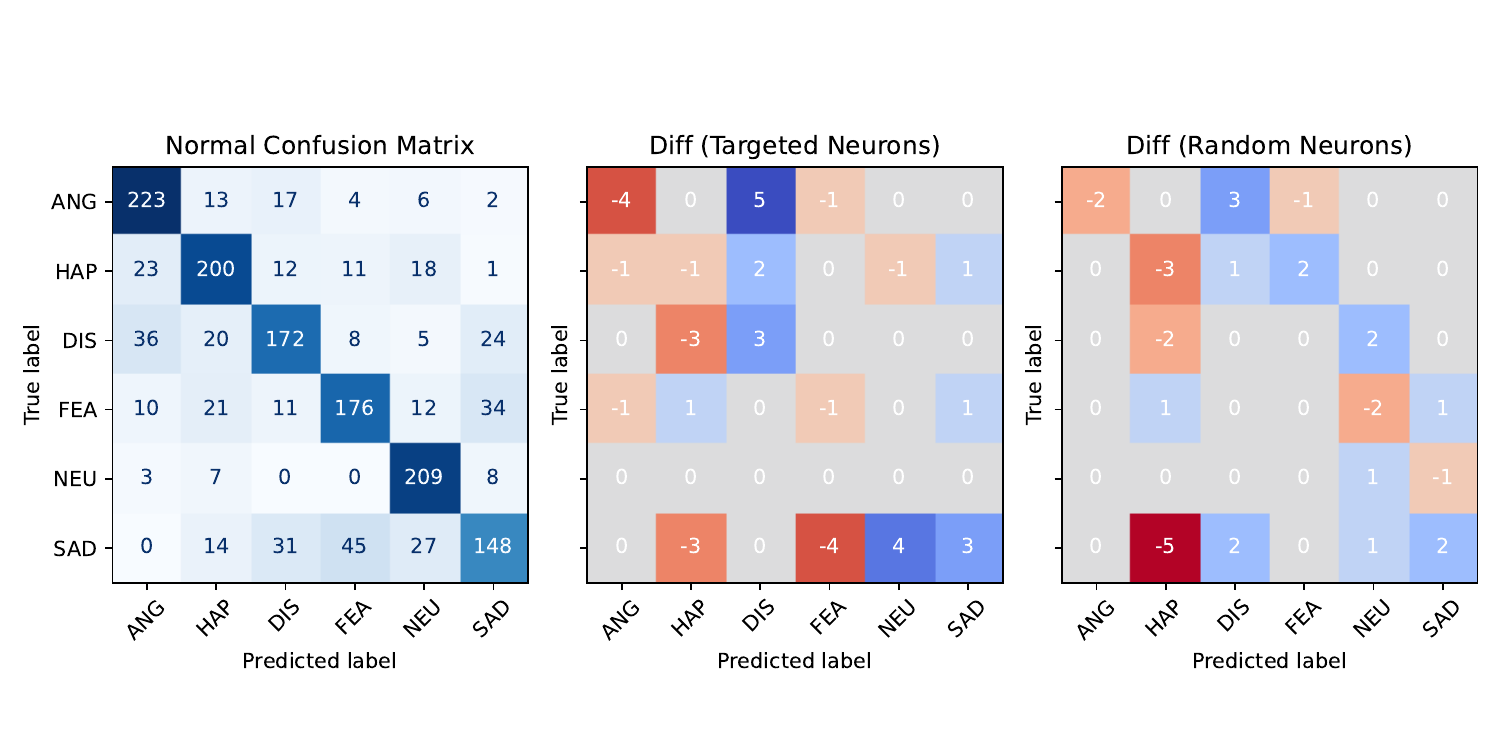}}
\vspace{-15pt}
\caption{Ablation impact on CREMA-D genre classification. Left: Original model. Middle/Right: Deviations after ablating class-specific vs. random neurons. The results indicate that class-specific neurons have a clearer targeted impact on classification compared to the random baseline.}
\label{fig-cm-emotion}
\end{figure}

In Fig.~\ref{fig-cm-emotion} (CREMA-D), deactivating targeted neurons degrades the accuracy of ``ANG'' and ``HAP,'' while other classes show minor fluctuations. In contrast, random ablation leads to smaller, scattered impacts across the classes.
These results suggest that the identified class-specific neurons contribute to the representation of their associated classes.


In summary, these observations validate our LAPE-based approach for identifying neurons and confirm that these class-specific neurons play a functional role in classification.

\section{Conclusion}
This paper investigated how general-purpose audio SSL models generalize to previously unseen tasks through the lens of mechanistic interpretability.
We showed that SSL neurons provide near-complete coverage of the classes defined in diverse unseen tasks and that these neurons have measurable effects on the classification performance of the classes to which they responded.
Notably, shared neuronal responses emerged for speech attributes such as gender, language family, and arousal-based emotional structure, as well as for the musical attribute of pitch octave classes, while neurons sensitive to acoustic similarity appeared to contribute to classes, such as music genres.

Our findings provide a neuron-level, mechanistic understanding of the generalization capability of general-purpose audio models, opening new directions for principled advances in the development of audio foundation models.

\bibliographystyle{IEEEtran}
\bibliography{ref}

\end{document}